\documentstyle[11pt,fewbody2005,twoside,epsfig,natbib,amsmath]{article}

\begin{document}

%%%%%%%%%%%%%%%%%%%%%%%%%%%%%%%%%%%%%%%%%%%%%%%%%%%%%%%%%%%%%%%%%%%%%%%%%%%%%%
\title{Star Cluster Evolution with Primordial Binaries}

%%%%%%%%%%%%%%%%%%%%%%%%%%%%%%%%%%%%%%%%%%%%%%%%%%%%%%%%%%%%%%%%%%%%%%%%%%%%%%
\author{John M. Fregeau${}^1$, M. Atakan G\"urkan${}^2$, \& Frederic A. Rasio${}^1$}
\affil{${}^1\,$Northwestern University, ${}^2\,$Sabanc\i{} University}

%%%%%%%%%%%%%%%%%%%%%%%%%%%%%%%%%%%%%%%%%%%%%%%%%%%%%%%%%%%%%%%%%%%%%%%%%%%%%%
\begin{abstract}
Observations and theoretical work suggest that globular clusters may be
born with initially very large binary fractions.  We present first results 
from our newly modified
Monte-Carlo cluster evolution code, which treats binary interactions exactly
via direct $N$-body integration.  It is shown that binary scattering interactions
generate significantly less energy than predicted by the recipes that
have been used in the past to model them in approximate cluster
evolution methods.  The new result that the cores of globular
clusters in the long-lived binary-burning phase are smaller than previously predicted
weakens the agreement with observations, thus implying that
more than simply stellar dynamics is at work in shaping the globular
clusters we observe today.
\end{abstract}

\keywords{stellar dynamics, methods: $N$-body simulations, globular clusters: general}

%%%%%%%%%%%%%%%%%%%%%%%%%%%%%%%%%%%%%%%%%%%%%%%%%%%%%%%%%%%%%%%%%%%%%%%%%%%%%%
\section{Introduction}

Observations, in combination with theoretical work, suggest that although the 
currently observed binary fractions in the cores of globular clusters 
may be small ($\sim 10\%$), the initial cluster binary fraction may have
been significantly larger ($\sim 100\%$) \citep{2005MNRAS.358..572I}.  It has been understood 
theoretically for many years that primordial binaries in star clusters act as an
energy source (through super-elastic scattering encounters with stars and other binaries), 
with a binary fraction of
a few percent being enough to postpone deep core collapse for many relaxation times 
\citep[see][for discussion and references]{2003ApJ...593..772F}.
In addition to playing a large part in the global evolution of globular clusters,
dynamical interactions of binaries also strongly affect the formation and evolution
of stellar and binary exotica (e.g.\ low-mass X-ray binaries, blue stragglers).

In previous studies using approximate methods like Monte-Carlo or Fokker-Planck
to simulate the evolution of star clusters, binary interactions had generally been 
treated using recipes culled from the results of large numbers of numerical scattering 
experiments---although in one case direct integration of binary interactions was performed 
for equal-mass stars \citep{2003MNRAS.343..781G}.  The recipes are typically known 
only for equal-mass binary interactions, thus prohibiting the use of a mass function 
in the cluster's initial conditions.  In order to model realistic clusters, which 
contain a wide range of masses, one must numerically integrate each binary interaction 
in order to resolve it properly.

In this article we present first results from our newly modified Monte-Carlo code, 
in which we have included for the first time exact integration of dynamical interactions 
of binaries with arbitrary mass members.  A detailed description of all new modifications (which
include physical stellar collisions and improvements to the core Monte-Carlo technique), and 
results, will be reported in a forthcoming paper.

%%%%%%%%%%%%%%%%%%%%%%%%%%%%%%%%%%%%%%%%%%%%%%%%%%%%%%%%%%%%%%%%%%%%%%%%%%%%%%%
\section{Results}

In a previous paper \citep{2003ApJ...593..772F}, we simulated the evolution
of clusters with primordial binaries by using recipes for the outcomes
of binary interactions.  Since we now treat binary interactions exactly
via direct numerical integration, we compare our results with those
of that work.

%%%%%%%%%%%%%%%%%%%%%%%%%%%%%%%%%%%%%%%%%%%%%%%%%%%%%%%%%%%%%%%%%%%%%%%%%%%%%%%%
\begin{figure}
  \begin{center}
    \includegraphics[width=0.7\textwidth]{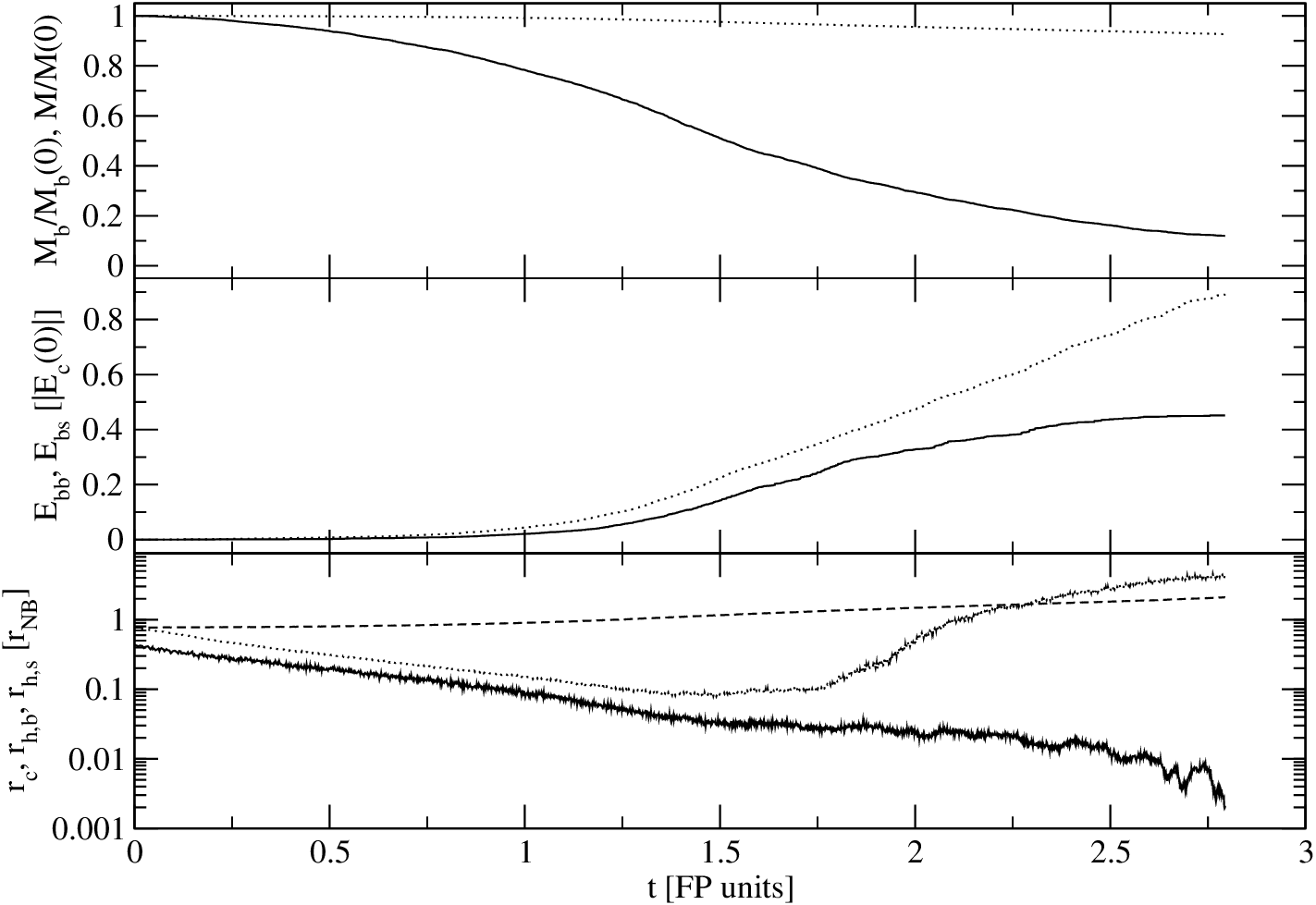}
    \caption{\small Evolution of an isolated Plummer model with $3\times 10^5$ stars and an 
      initial 2\% binary fraction.  The top panel shows $M_b$, 
      the total mass in binaries bound to the cluster (solid line), and $M$, the total mass 
      of the cluster (dotted line), as a function of time, relative to their initial values.  
      The middle panel shows $E_{\rm bb}$, the cumulative energy generated in binary--binary 
      interactions (solid line) and $E_{\rm bs}$, the cumulative energy generated in binary--single 
      interactions (dotted line) relative to $|E_c(0)|$, the absolute value of the cluster's 
      initial mechanical energy.  The bottom panel shows the evolution of $r_c$, the cluster 
      core radius (solid line), $r_{\rm h,b}$, the half-mass radius of the binaries (dotted line), 
      and $r_{\rm h,s}$, the half-mass radius of single stars (dashed line).  The unit of time is 
      the global cluster relaxation time, which is $\approx 10 t_{\rm rh}$.\label{fig:pl_n3e5_fb0.02.binary}}
  \end{center}
\end{figure}
%%%%%%%%%%%%%%%%%%%%%%%%%%%%%%%%%%%%%%%%%%%%%%%%%%%%%%%%%%%%%%%%%%%%%%%%%%%%%%%%

%%%%%%%%%%%%%%%%%%%%%%%%%%%%%%%%%%%%%%%%%%%%%%%%%%%%%%%%%%%%%%%%%%%%%%%%%%%%%%%%
\begin{figure}
  \begin{center}
    \includegraphics[width=0.7\textwidth]{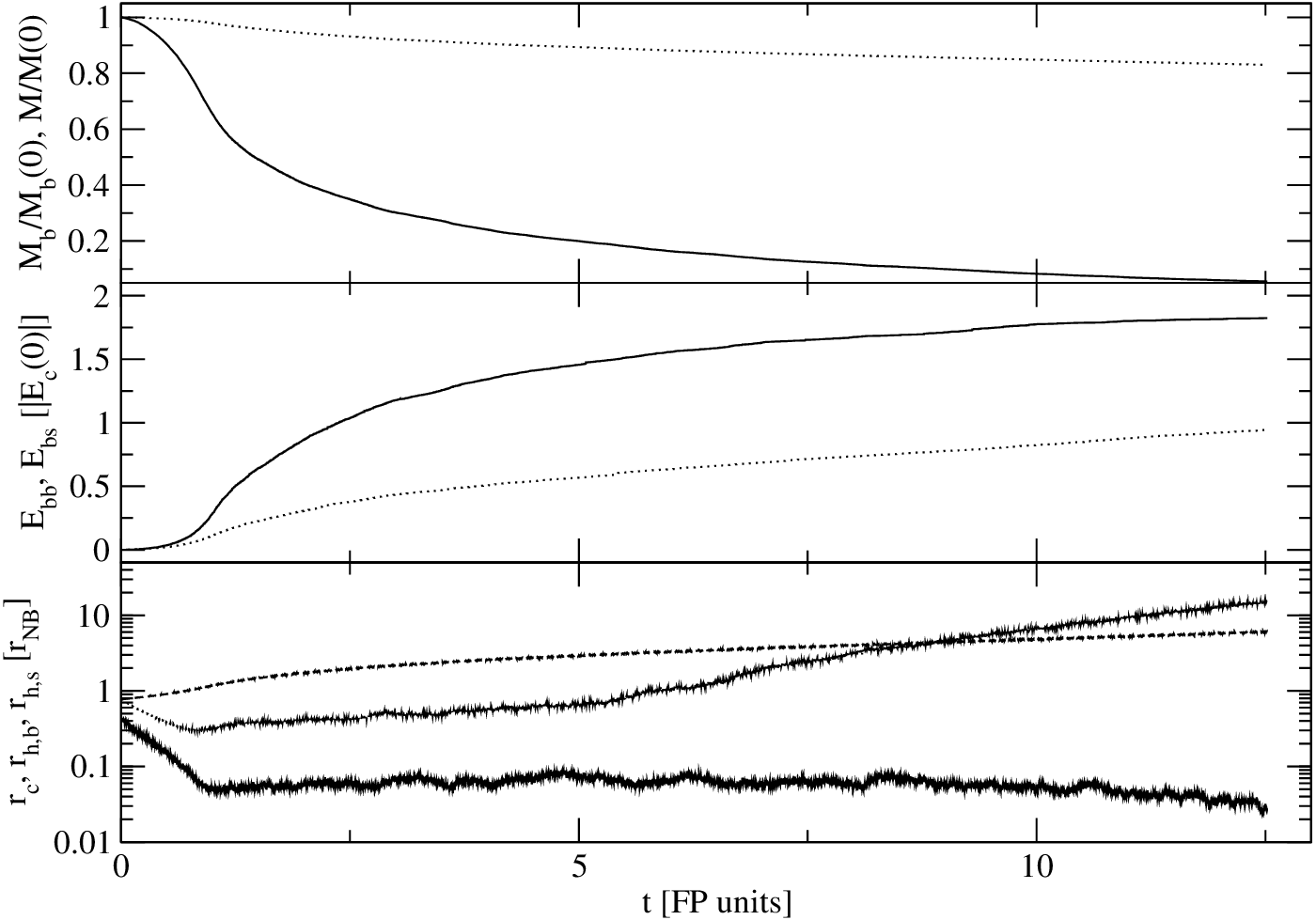}
    \caption{\small Same as Fig.~\ref{fig:pl_n3e5_fb0.02.binary}, but for an initial 
      binary fraction of 10\%.\label{fig:pl_n3e5_fb0.1.binary}}
  \end{center}
\end{figure}
%%%%%%%%%%%%%%%%%%%%%%%%%%%%%%%%%%%%%%%%%%%%%%%%%%%%%%%%%%%%%%%%%%%%%%%%%%%%%%%%

%%%%%%%%%%%%%%%%%%%%%%%%%%%%%%%%%%%%%%%%%%%%%%%%%%%%%%%%%%%%%%%%%%%%%%%%%%%%%%%%
\begin{figure}
  \begin{center}
    \includegraphics[width=0.7\textwidth]{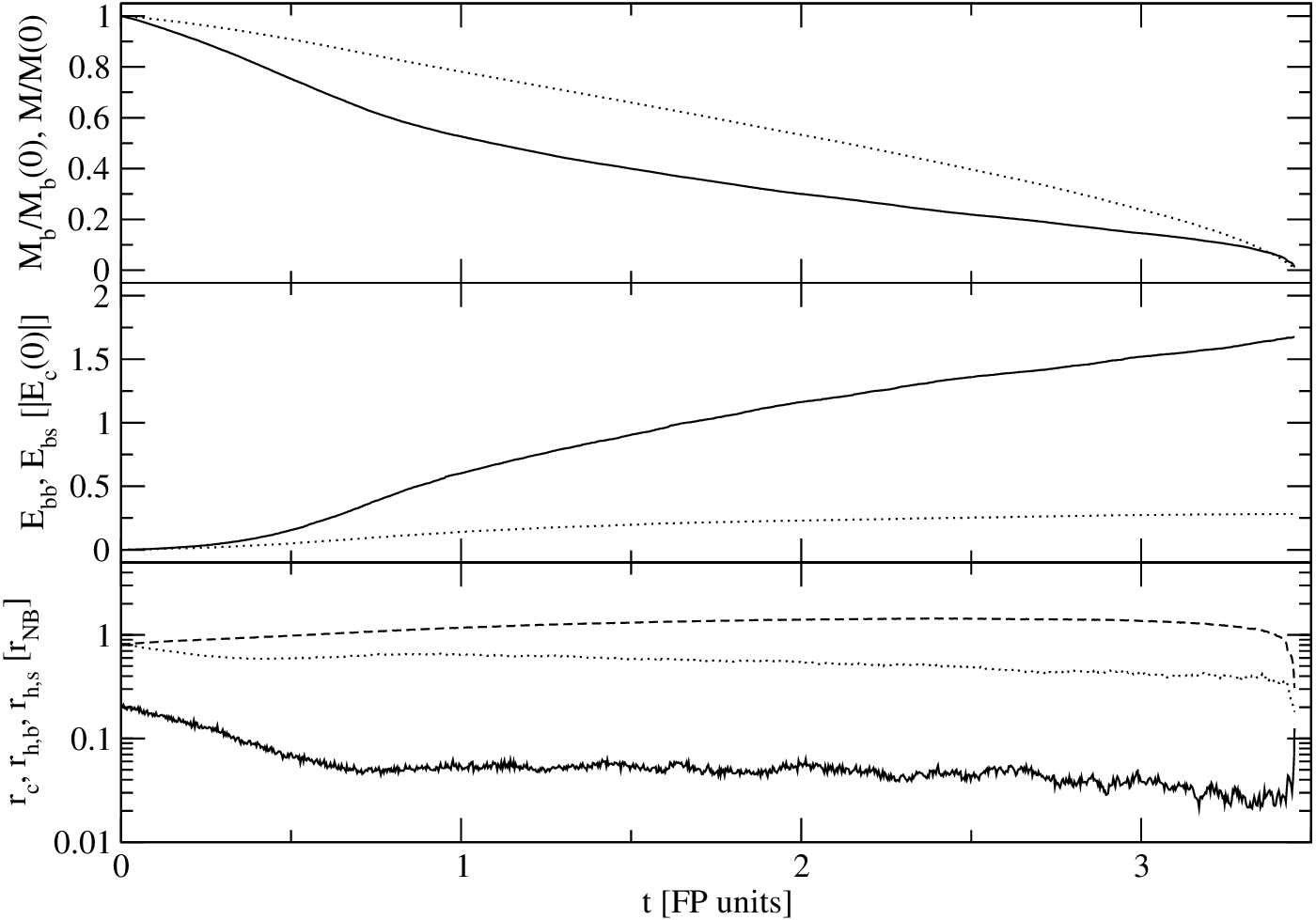}
    \caption{\small Same as Fig.~\ref{fig:pl_n3e5_fb0.02.binary}, but for a tidally-truncated
      $W_0=7$ King model with an initial binary fraction of 20\%.\label{fig:king_w7_n3e5_fb0.2.binary}}
  \end{center}
\end{figure}
%%%%%%%%%%%%%%%%%%%%%%%%%%%%%%%%%%%%%%%%%%%%%%%%%%%%%%%%%%%%%%%%%%%%%%%%%%%%%%%%

Fig.~\ref{fig:pl_n3e5_fb0.02.binary} shows the evolution of an isolated Plummer model with 
an initial 2\% binary fraction.  In 
this model all stars had the same mass.  This can be compared directly with
Fig.~3 of \citet{2003ApJ...593..772F}, in which we used recipes for binary interactions.  
The evolution of the two models is 
quite similar, with both
showing the binaries starting to get kicked out of the core (as evidenced by $r_{\rm h,b}$)
around $t/t_{\rm rh}\approx 17$, entering deep core collapse at $t/t_{\rm rh}\sim 25$,
showing roughly the same rate of binary mass loss (due to binaries getting disrupted or
kicked out of the cluster), and roughly the same rate of energy generation in binary interactions.
However, we find that although the simple recipes reproduce the energy generation in binary--single 
interactions reasonably well, they overestimate the rate of energy generation in binary--binary
interactions by a factor of two relative to exact integrations.  The same disagreement between
recipes and exact integrations for binary--binary interactions was
found by \citet{2003MNRAS.343..781G} in their models.  Another point of note is that 
with direct integrations the approach to deep core collapse is much noisier, with what 
appear to be many ``mini'' gravothermal oscillations.

Fig.~\ref{fig:pl_n3e5_fb0.1.binary} shows the same as Fig.~\ref{fig:pl_n3e5_fb0.02.binary}, but
for an initial binary fraction of 10\%.  This can be compared directly with Fig.~4 of
\citet{2003ApJ...593..772F}.  Again we find that the energy generation rate due to binary--single
interactions is comparable between recipes and direct integrations.  For binary--binary,
integrations predict a rate that is roughly 60\% of what recipes predict.

Finally, in Fig.~\ref{fig:king_w7_n3e5_fb0.2.binary}
we show the evolution of an initial $W_0=7$ King model with 20\% primordial
binaries, which can be compared directly with Fig.~11 of \citet{2003ApJ...593..772F}.
Just as in that paper, the cluster is disrupted by the tidal field of its host galaxy
before core collapse, at a time of roughly $t/t_{\rm rh}\sim 35$--$40$.  Looking
more closely at the figures, we see that the evolution of $M_b$, $M$, $r_{\rm h,s}$,
and $r_{\rm h,b}$ is quite similar between the two models.  Again, we find that
with integrations the binary-burning rate is smaller.  One major difference between
the two is that with integrations the core radius shrinks, then remains at a constant
value once the steady binary-burning phase is reached.  With recipes, the core appears
to expand from the start.  The discrepancy is no doubt caused by the larger
rate of binary-burning in the model with recipes.

%%%%%%%%%%%%%%%%%%%%%%%%%%%%%%%%%%%%%%%%%%%%%%%%%%%%%%%%%%%%%%%%%%%%%%%%%%%%%%%
\section{Conclusion}

Although our models for evolving globular clusters are somewhat simplified---since
they include the effects of two-body relaxation and binary interactions, but ignore
the effects of stellar evolution---we can still reach some conclusions
by comparing with observations.
In \citet{2003ApJ...593..772F} we found that the values of cluster half-mass radius
to core radius, $r_h/r_c$, predicted by models using recipes for binary interactions
were in reasonably good agreement with observations.  Our new models, which treat
binary interactions exactly, predict a significantly lower core energy generation rate,
and a consequently smaller core, yielding values of $r_h/r_c$ up to an order of
magnitude larger.  This disagreement
suggests that processes other than simply stellar dynamics---such as stellar
evolution---may play a very important role in shaping the clusters 
we observe today.

%%%%%%%%%%%%%%%%%%%%%%%%%%%%%%%%%%%%%%%%%%%%%%%%%%%%%%%%%%%%%%%%%%%%%%%%%%%%%%
\acknowledgments This work was supported by NASA Grant NAG5-12044.  
JMF thanks the conference organizers and the University of Turku
for their hospitality during the Few-Body conference.

%%%%%%%%%%%%%%%%%%%%%%%%%%%%%%%%%%%%%%%%%%%%%%%%%%%%%%%%%%%%%%%%%%%%%%%%%%%%%%%
\bibliographystyle{apj}
\bibliography{apj-jour,main}

\end{document}